\documentclass[12pt]{article}

\usepackage{amsmath,amssymb,amsfonts}
\usepackage{graphicx}
\usepackage{xcolor}
\usepackage{xr-hyper}   
\usepackage{hyperref}

\makeatletter
\newcommand*{\addFileDependency}[1]{%
  \typeout{(#1)}%
  \@addtofilelist{#1}%
  \IfFileExists{#1}{}{\typeout{No file #1.}}%
}
\makeatother
\newcommand*{\myexternaldocument}[2][]{%
  \externaldocument[#1]{#2}%
  \addFileDependency{#2.tex}%
  \addFileDependency{#2.aux}%
}

\myexternaldocument[]{supplementary}
\begin{document}

\title{Instability-induced bistable shape-morphing kirigami structures}

\author{
Xiaoyuan Ying$^{1,*}$ and Marcelo A. Dias$^{1,*}$\\[6pt]
\small $^{1}$Institute for Infrastructure and Environment, University of Edinburgh,\\
\small Thomas Bayes Road, Edinburgh EH9 3FG, United Kingdom\\
\small $^{*}$Corresponding author: Xiaoyuan.Ying@ed.ac.uk, Marcelo.dias@ed.ac.uk
}

\date{}

\maketitle{}

\begin{abstract}
Deployable shape-morphing structures that transform from flat sheets into stable three-dimensional configurations are highly desirable for applications ranging from soft robotics and biomedical devices to adaptive architecture and aerospace systems. Existing kirigami-based morphing systems primarily rely on isotropic deployment, compliant soft materials, or external constraints to maintain deployed shapes, which limits geometric programmability, structural integrity, and applicability in rigid-material systems. Here, we present an inverse design framework for anisotropic bistable kirigami structures that enables programmable shape morphing through controlled geometric frustration and instability-induced deployment. The framework combines a semi-analytical mechanical model with geometry to establish a direct connection between geometric transformation and the underlying energy landscape. We show that instability-induced shape morphing leads to tunable bistability and directional deployment in anisotropic kirigami structures. The results are validated through finite element simulations and experiments, demonstrating stable deployed configurations and programmable anisotropic morphing. The proposed framework further provides a general design strategy that can be integrated with various active actuation systems, enabling broader engineering applications.
\end{abstract}

\vspace{0.5em}

\noindent\textbf{Keywords:} Bistable kirigami; anisotropic shape morphing; geometric frustration; instability-induced; inverse design

\vspace{1cm}

\section{Introduction}
\label{Introduction}
Shape-morphing structures are engineered systems capable of transitioning between distinct configurations in response to external stimuli or through intrinsic and self-regulated mechanisms, enabling programmable changes in geometry, functionality, and mechanical response. Such transformations can be triggered by mechanical loading~\cite{hwang2022shape,liu2023snap,taffetani2018static,rafsanjani2016bistable,zaman2025one,panetta2021computational}, thermal fields~\cite{ge2016multimaterial,zang20244d,tang2019programmable}, swelling~\cite{kim2012designing,dias2011programmed,zhou2018shape,su20184d}, magnetic actuation~\cite{wang2023physics,alapan2020reprogrammable,chi2024magnetic}, electric stimuli~\cite{jeon2025super,xu2025electric}, or active constituents within the material~\cite{keber2014topology,salbreux2017mechanics}. Among these systems, mechanical metamaterials~\cite{bertoldi2017flexible,surjadi2019mechanical} achieve unconventional functionalities primarily through architectural design rather than material composition, providing a powerful platform for programmable shape morphing across multiple length scales.

Kirigami-based mechanical metamaterials, formed by introducing patterned cuts into thin sheets, offer a particularly effective route for large and reversible geometric transformations~\cite{hong2022boundary,ying2025inverse}. Compared with origami systems that rely primarily on fold kinematics~\cite{chen2022origami,walker2026spherical}, kirigami structures enable larger deformation through the combined effects of rotation, bending, stretching, and elastic instability~\cite{sadik2021local,sadik2022local}. These characteristics make kirigami highly attractive for applications in deployable structures, biomedical devices, and soft robotics. In particular, instability-induced shape morphing, in which elastic buckling and snap-through are harnessed as functional design mechanisms rather than treated as failure modes, enables bistability and self-locking deployment without continuous external actuation~\cite{reis2015perspective,hu2015buckling}. A familiar example is the spring-loaded light switch, which possesses two stable states (on and off) and transitions between them under a single push. Similarly, the snap-through instability of a clamped--clamped beam can be exploited to generate robotic motion~\cite{chen2018harnessing}, and spherical caps~\cite{taffetani2018static} can undergo large shape change with rapid energy release, passing through the intermediate configurations without residing in them.

Most existing kirigami morphing systems are monostable~\cite{konakovic2016beyond,konakovic2018computational}, where the deployed configuration is not an energy minimum and therefore requires sustained actuation, boundary constraints, or auxiliary locking mechanisms to maintain. Recent studies~\cite{chen2021bistable,shang2018durable} have addressed this limitation by introducing bistable unit cells that allow both undeployed and deployed states to remain mechanically stable. However, many of these systems rely on highly compliant soft materials, such as elastomers or rubber-like substrates, where the deployed state is often governed by compliant deformation, viscoelastic effects, or geometric locking rather than a sharply defined mechanical energy barrier. As a result, the bistable response is often weak, and the deployed configuration may behave more like a compliant mechanism than a robust rigid-material bistable structure. In addition, these approaches are largely restricted to isotropic deployment, where uniform scaling is assumed across the structure ~\cite{panetta2021computational, chen2021bistable,konakovic2018rapid}. Such assumptions limit geometric programmability and fail to capture anisotropic deformation, where directional stretching fundamentally alters geometric compatibility, redistributes elastic energy, and strongly influences bistability.

Here, we present a mechanics-guided inverse design framework for rigid bistable kirigami tessellations. Unlike conventional deployable structures that rely on kinematic mechanisms such as hinges, linkages, and foldable joints, where stable deployment often requires additional locking systems or continuous actuation, our approach exploits instability-induced bistability arising from elastic deformation of slender ligaments connecting quarigid units. This enables monolithic fabrication, self-locking deployment, and improved structural stiffness in rigid-material systems.

The proposed framework combines a semi-analytical model based on a discretized Hencky bar-chain formulation with differential geometry to establish a direct connection between anisotropic geometric transformation and the underlying energy landscape. By explicitly incorporating anisotropic deformation into the inverse design pipeline, we construct a unit library that links geometric anisotropy to bistable strain and stability metrics, enabling mechanics-consistent design of deployable tessellated surfaces. Finite element simulations and experiments on laser-cut prototypes validate the predicted deployment pathways and stable deployed configurations. Our results provide a general strategy for designing rigid-material bistable metamaterials with programmable anisotropic deployment and broaden the applicability of kirigami-based shape morphing to load-bearing and reusable engineering systems.

\section{Results}
\label{chapter 2}
\subsection{Characterization and computational model of unit cell}
\begin{figure}[!ht]
    \centering
    \includegraphics[width=0.93\linewidth]{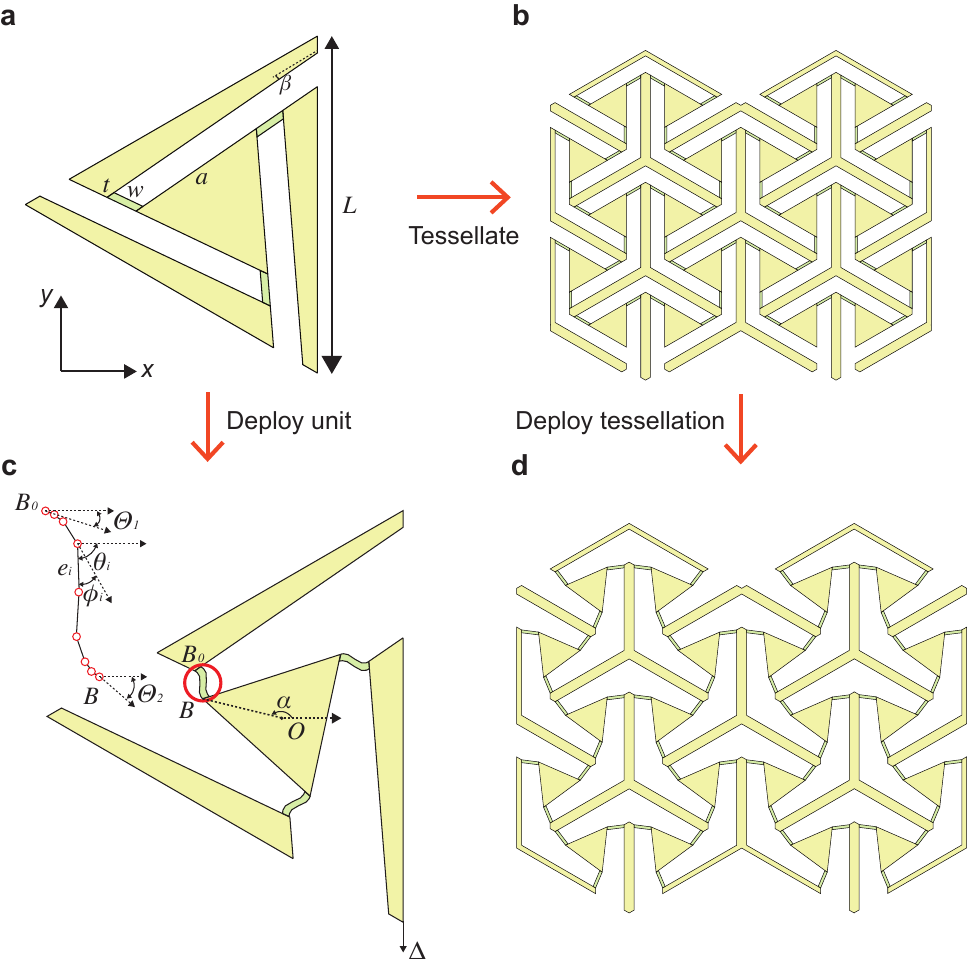}
\caption{\textbf{Kirigami unit cell and tessellation.}
(\textbf{a})~Triangular auxetic unit cell parametrised by ligament
thickness $t$, gap width $w$, in-radius $a$, tilting angle $\beta$,
and unit cell length $L$.
(\textbf{b})~Undeployed $4\times5$ tessellation.
(\textbf{c})~Deployed unit cell with discrete beam representation of
one ligament: nodes (circles) connected by extensible bars with
segment angles $\theta_i$, relative rotations $\varphi_i$, and axial
extensions $e_i$; $\Theta_1$, $\Theta_2$ are the prescribed end
tangent angles; $\Delta$ is the applied end displacement.
(\textbf{d})~Deployed tessellation.}
\label{fig:tessellation}
\end{figure}

We focus on triangular auxetic units formed by introducing
three mutually intersecting slits into an equilateral triangular
sheet~\cite{rafsanjani2016bistable}. This cut pattern defines three
slender elastic ligaments that connect a triangular core to
the surrounding flanks, forming the fundamental building block of
the bistable kirigami structure (Fig.~\ref{fig:tessellation}a).
The unit geometry is fully parameterised by five quantities: the
unit cell length $L$, ligament thickness $t$, gap width $w$,
tilting angle $\beta$, and triangular core in-radius $a$; all remaining
geometric quantities follow from these through trigonometric
relations. Periodic arrangement over a triangular lattice yields the
full kirigami tessellation (Fig.~\ref{fig:tessellation}b).

Existing deployable auxetic designs rely on one of two idealised
connector models. Point-wise hinge mechanisms~\cite{%
friedrich2017locally,konakovic2018computational} set $w\to 0$ and
treat bistability as a purely kinematic phenomenon. Finite-width hinge
models~\cite{chen2021bistable,kim2025design} recover material
dependence but obtaining accurate energy profiles typically required finite element analysis. Our design replaces both idealisations with slender elastic
ligaments capable of large coupled bending--stretching deformation,
enabling monolithic fabrication from rigid sheet materials, improved
fatigue resistance~\cite{shang2018durable}, and the tractable
semi-analytical formulation developed below.

Two simplifying assumptions make the analysis tractable. First, the
flanks and triangular core are treated as rigid bodies, since their
compliance is negligible relative to that of the slender ligaments.
Second, ligament--flank and ligament--core interfaces are idealised
as perfect rotational joints, permitting end rotation without local
stress concentration.

Each ligament is modelled using the Hencky--bar--chain (HBM)
formulation~\cite{hencky1921angenaherte}, which discretises the
continuous Euler--Bernoulli beam into $N$ extensible bars connected
by $K = N-1$ rotational springs(Fig.~\ref{fig:tessellation}c). The
unknown state vector is
\begin{equation}
\mathbf{x}
=
\bigl[\varphi_1,\ldots,\varphi_K,\;
e_1,\ldots,e_N,\;\alpha\bigr]^{\!\mathsf{T}},
\label{eq:dof}
\end{equation}
where $\varphi_i$ is the relative rotation between segments $i$ and
$i{+}1$, $e_i$ is the axial extension of segment $i$, and $\alpha$
is the rigid-body rotation of the inner triangular core. The
discrete elastic energy is
\begin{equation}
E_{\mathrm{lig}}(\mathbf{x})
=
\frac{1}{2}\sum_{i=1}^{K} k_{b,i}\,\varphi_i^{2}
+
\frac{1}{2}\sum_{i=1}^{N} k_{s,i}\,e_i^{2},
\label{eq:energy}
\end{equation}
where $k_{b,i} = EI/h_i$ and $k_{s,i} = EA/l_i$ are the discrete
bending and axial stiffnesses, with $h_i = \tfrac{1}{2}(\ell_i +
\ell_{i+1})$. This energy converges to the FEA result
as $N\to\infty$; however, the rate of convergence depends critically
on the discretisation strategy (Supplementary ~\ref{sec:S_HBM}),
an end-clustered discretisation achieves
a relative error within $5\%$ of full FEM using only $N = 12$
segments (fig. ~\ref{fig:S_convergence}).

The equilibrium configuration is obtained by solving the constrained
minimisation problem (Supplementary ~\ref{sec:S_isotropic}), which is given by
\begin{equation}
\min_{\mathbf{x}}\; E_{\mathrm{lig}}(\mathbf{x})
\qquad \text{subject to} \qquad
g_{\Theta}(\mathbf{x}) = 0, \quad
\mathbf{g}_{r}(\mathbf{x}) = \mathbf{0},
\label{eq:opt}
\end{equation}
where the two constraints enforce geometric compatibility at the
ligament ends. The \emph{angle-sum constraint}
\begin{equation}
g_{\Theta}(\mathbf{x})
=
\Theta_1 + \sum_{i=1}^{K}\varphi_i - \Theta_2 = 0
\label{eq:angle}
\end{equation}
requires the cumulative segment rotations to match the prescribed
end tangent angles $\Theta_1$ and $\Theta_2$. The
\emph{positional closure constraint}
\begin{equation}
\mathbf{g}_{r}(\mathbf{x})
=
\mathbf{B}_0
+ \sum_{i=1}^{N}(\ell_i + e_i)
\begin{bmatrix}\cos\theta_i \\ \sin\theta_i\end{bmatrix}
- \mathbf{B} = \mathbf{0}
\label{eq:closure}
\end{equation}
requires the deformed ligament to connect its two prescribed
end-points $\mathbf{B}_0$ and $\mathbf{B}$, where the cumulative
segment angle is $\theta_i = \Theta_1 + \sum_{j=1}^{i-1}\varphi_j$.
This constraint minimisation, defined by eqs.~\eqref{eq:opt}--\eqref{eq:closure}, is solved using the
interior-point method (\texttt{fmincon}, MATLAB).

\begin{figure}[!ht]
    \centering
    \includegraphics[width=0.97\linewidth]{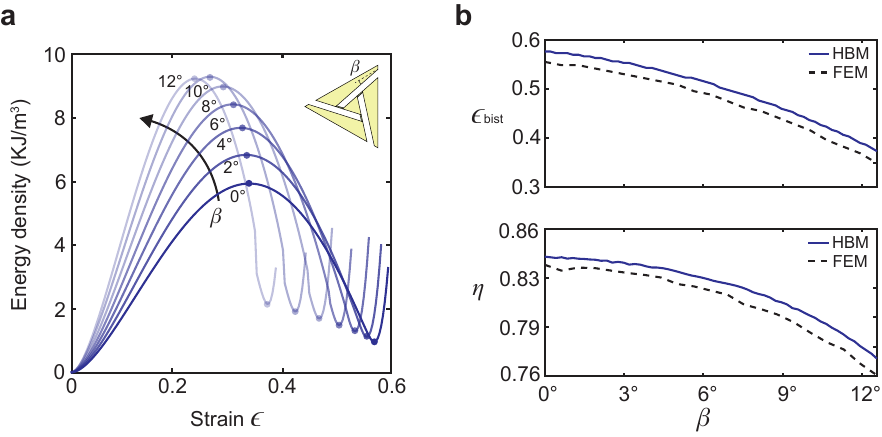}
    \caption{\textbf{Bistability curves and Comparison between HBM and FEM.}
    (a)~Strain energy density versus strain for units with different tilting angle $\beta$ computed with the semi-analytical model.
    (b)~Bistable strain $\epsilon_\text{bist}$(top) and bistability $\eta$(bottom) versus $\beta$ from HBM and FEM.Definitions of
    $\epsilon_\text{bist}$ and $\eta$ are given in Supplementary Fig.~\ref{fig:deployment_energy}.}
    \label{fig:sweep_study}
\end{figure}

Figure~\ref{fig:sweep_study}(a) shows the strain-energy density of a single
unit during deployment, computed with our Hencky bar-chain model, for tilting
angles $\beta$ from $0^\circ$ to $9^\circ$. Each curve exhibits bistable behaviour: an energy barrier at intermediate strain followed by a second local minimum at the deployed state. $\beta$ acts
as a direct geometric handle on both the height and the location of the energy
landscape. The two scalar metric used throughout are defined in
Supplementary Fig.~\ref{fig:deployment_energy}: the bistable strain
$\epsilon_\mathrm{bist}$ is the strain at which the deployed-state energy
minimum occurs, and the bistability
$\eta = \Delta U/U_{\max} = (E_{\max}-E_{\min})/E_{\max}$ is the normalised
depth of that minimum relative to the energy barrier.

Figure~\ref{fig:sweep_study}(b) compares these two metrics from HBM and FEM
over the same range. The two methods agree well overall: the HBM overestimates
$\epsilon_\text{bist}$ by about 2\% and $\eta$ by about 1\%, confirming it as
an accurate and more efficient tool calculation.

\begin{figure}[!ht]
    \centering
    \includegraphics[width=0.95\linewidth]{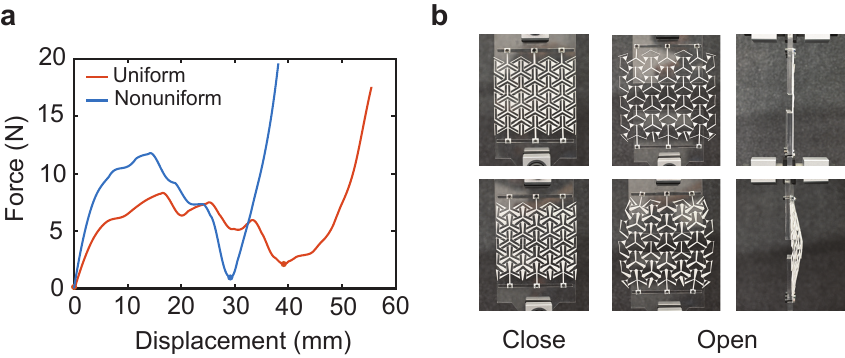}
     \caption{\textbf{Uniaxial tensile test of uniform and nonuniform tessellations.}
    (a)~force-displacement curve during loading(tension).
    (b)~Photographs of uniform tessellation(top) and nonuniform tessellation(bottom) in the closed state and open state with side views.}
    \label{fig:tessellation_test}
\end{figure}

To examine how the unit-level bistability carries over to the tessellation scale, Figure~\ref{fig:tessellation_test} compares the uniaxial tensile response of two
fabricated specimens: a uniform tessellation of identical units and a nonuniform
tessellation whose units have spatially varying tilting angle $\beta$. Both force--displacement
curves exhibit snap-through, marked by a force drop as the units deploy, and both
specimens remain stable in the open state once unloaded (Fig.~\ref{fig:tessellation_test}b). The two cases reach snap-through at different displacements
because their constituent units have different bistable strains $\epsilon_\mathrm{bist}$.
In the nonuniform tessellation, this strain mismatch prevents adjacent units from remaining
in-plane, producing geometric frustration: the specimen develops a curved boundary and
pronounced out-of-plane deformation during stretching, whereas the uniform tessellation
deploys essentially in-plane.

\subsection{Effect of anisotropic deployment and geometric 
design on bistability}
\label{sec:bistability}
In a tessellated structure, geometric compatibility requires adjacent 
units to share edges while accommodating spatially varying stretch. 
Previous studies~\cite{konakovic2018computational,chen2021bistable,%
kim2025design} commonly assume isotropic expansion---that all edges 
of a unit scale uniformly---which greatly simplifies analysis but 
neglects the anisotropic deformation that is unavoidable in any 
non-trivially shaped tessellation. We show that such anisotropic 
deformation can fundamentally alter the bistable energy landscape, 
rendering the isotropic assumption insufficient for reliable bistable 
design(Supplementary ~\ref{sec:S_anisotropic}).

\begin{figure}[!ht]
    \centering
    \includegraphics[width=0.93\linewidth]{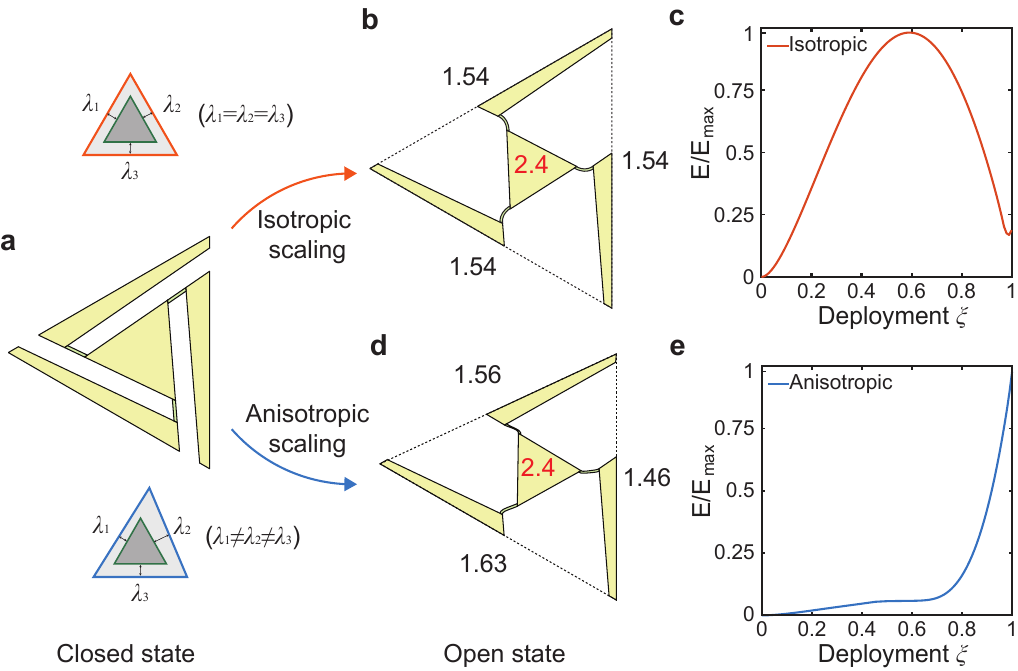}
    \caption{\textbf{Effect of deployment path on bistability.}
    (\textbf{a})~Initial closed state.
    (\textbf{b})~Isotropically scaled open configuration with uniform
    edge scale factors $(\lambda_1,\lambda_2,\lambda_3) = (1.54, 1.54, 1.54)$.
    (\textbf{c})~Energy--displacement curve for isotropic scaling,
    exhibiting a well-defined bistable response.
    (\textbf{d})~Anisotropically scaled open configuration with unequal
    edge scale factors $(\lambda_1,\lambda_2,\lambda_3) = (1.46, 1.56, 1.63)$.
    (\textbf{e})~Energy--displacement curve for anisotropic scaling,
    yielding a monostable profile despite identical area expansion.}
    \label{fig:anisotropic_deployment}
\end{figure}

To parametrise the deployed configuration of the unit, we 
introduce edge-wise scale factors $(\lambda_1, \lambda_2, 
\lambda_3)$, defined as the ratio of the deformed to 
undeformed edge length,
\begin{equation}
    \lambda_i = \frac{\tilde{l}_i}{l_i}, \qquad i = 1,2,3.
\end{equation}
The isotropic case corresponds to $\lambda_1 = \lambda_2 = 
\lambda_3$, while anisotropic deployment is characterised by 
unequal scale factors. To evaluate the energy along a given 
deployment path, we introduce the scalar parameter $\xi \in 
[0,1]$, where $\mathbf{q}(0)$ and $\mathbf{q}(1)$ denote the 
initial and fully deployed nodal coordinates, respectively. 
The intermediate configuration is obtained by linear 
interpolation,
\begin{equation}
    \mathbf{q}(\xi) = (1-\xi)\,\mathbf{q}(0) 
    + \xi\,\mathbf{q}(1),
    \label{eq:linear_interpolation}
\end{equation}
which is physically motivated by the assumption that 
rigid-body rotations of the flanks and inner core vary 
smoothly and monotonically between the two states. The 
computational procedure for evaluating the energy along an 
anisotropic deployment path prescribed by 
$(\lambda_1, \lambda_2, \lambda_3)$.

Figure~\ref{fig:anisotropic_deployment} compares two 
deployment paths for the same unit geometry, both with an 
identical area scaling ratio of $2.4$. The first follows 
isotropic scaling, transitioning from the closed state 
(Fig.~\ref{fig:anisotropic_deployment}a) to a uniformly 
stretched configuration 
(Fig.~\ref{fig:anisotropic_deployment}b) with edge scale 
factors $(\lambda_1, \lambda_2, \lambda_3) = (1.54, 1.54, 1.54)$. 
The second follows anisotropic scaling to a non-uniformly 
stretched configuration 
(Fig.~\ref{fig:anisotropic_deployment}d) with edge scale 
factors $(\lambda_1, \lambda_2, \lambda_3) = (1.46, 1.56, 
1.63)$. Despite identical overall area expansion, the two 
paths produce fundamentally different energy landscapes 
(Fig.~\ref{fig:anisotropic_deployment}c,e): the isotropic 
path yields a pronounced bistable response ($\eta > 0$) with 
a well-defined energy well of depth $\Delta E$, whereas the 
anisotropic path produces a monostable profile with no local 
energy minimum. This demonstrates that 
bistability depends critically on the prescribed deployment 
path and the geometric parameters of the unit, motivating an 
explicit characterisation of how $\eta$ and 
$\epsilon_\text{bist}$ vary across the admissible anisotropic 
deformation space.

\begin{figure}[ht!]
    \centering
    \includegraphics[width=1\linewidth]{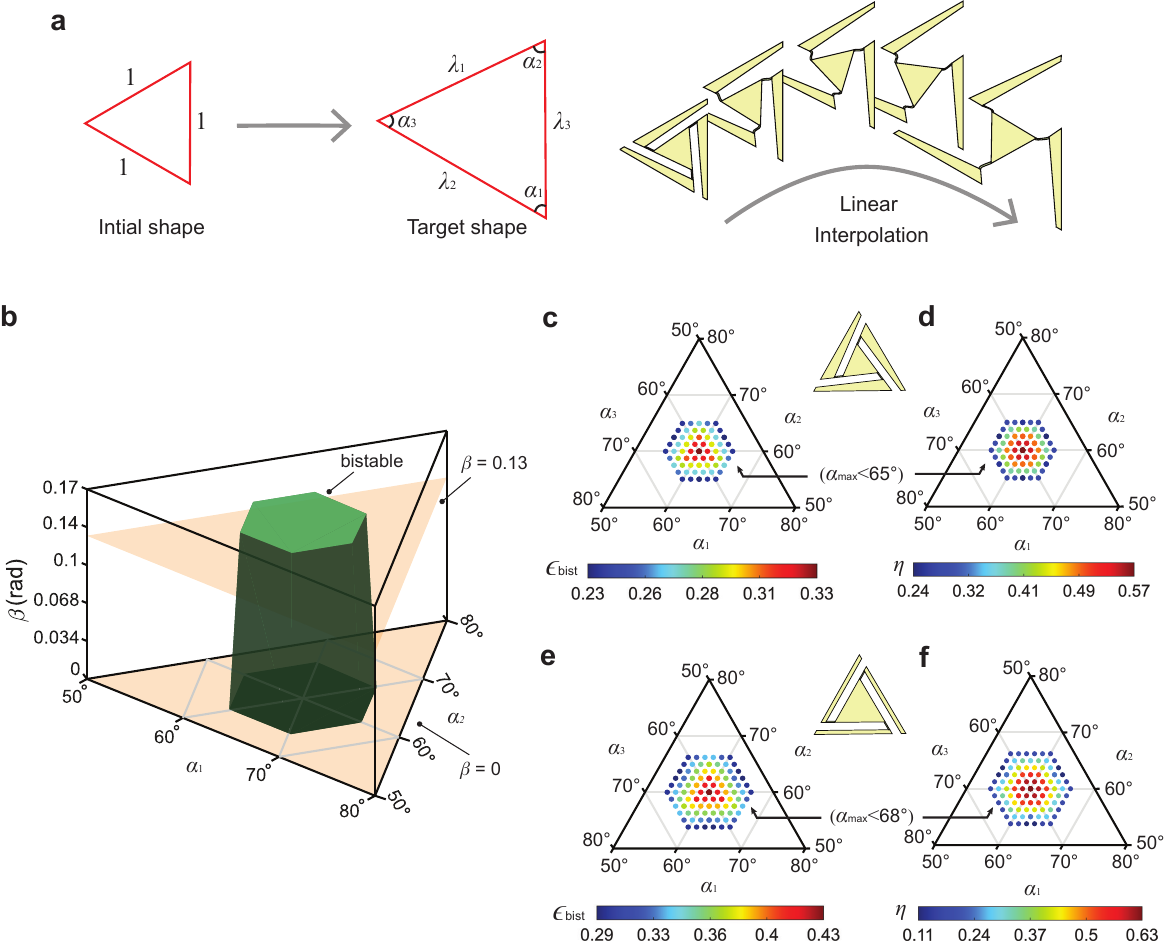}
    \caption{\textbf{Sensitivity of bistability to anisotropic
    deformation and tilting angle $\beta$.}
    (\textbf{a})~Each unit deforms from an equilateral initial shape
    to a target shape parametrised by edge scale factors
    $(\lambda_1, \lambda_2, \lambda_3)$ and internal angles
    $(\alpha_1, \alpha_2, \alpha_3)$; intermediate states follow
    linear interpolation of nodal coordinates.
    (\textbf{b})~Bistable region in $(\alpha_1, \alpha_2, \alpha_3,\beta)$
    space; highlighted planes correspond to the slices in (c--f).
    (\textbf{c,d})~Ternary maps at $\beta = 0.13$~rad
    ($\alpha_{\max} < 65^\circ$) showing $\epsilon_{\mathrm{bist}}$
    and $\eta$.
    (\textbf{e,f})~Ternary maps at $\beta = 0$
    ($\alpha_{\max} < 68^\circ$) showing $\epsilon_{\mathrm{bist}}$
    and $\eta$.}
    \label{fig:ternary_plot}
\end{figure}

We next examine how the tilting angle $\beta$ and anisotropic 
deformation jointly govern $\eta$ and $\epsilon_\text{bist}$.
As illustrated in Fig.~\ref{fig:ternary_plot}(a), each
triangular unit deforms from an equilateral initial shape
to a target shape characterised by edge scale factors
$(\lambda_1, \lambda_2, \lambda_3)$. Since the goal is for
the target shape to coincide with the bistable equilibrium
of the unit, we use $\epsilon_\text{bist}$ as the scalar
deployment strain. The anisotropic deployed state is then
parametrised equivalently by the internal angles
$(\alpha_1, \alpha_2, \alpha_3)$ of the deformed triangle,
ordered as $\alpha_1 \geq \alpha_2 \geq \alpha_3$ with
$\alpha_1 + \alpha_2 + \alpha_3 = 180^\circ$, as
\begin{equation}
    (\lambda_1,\,\lambda_2,\,\lambda_3)
    =
    \left(\frac{\sin\alpha_1}{\sin\alpha_3},\;
    \frac{\sin\alpha_2}{\sin\alpha_3},\;
    1\right)(1 + \epsilon_\text{bist}),
\end{equation}
reducing the three-dimensional scale-factor space to a
two-dimensional space of angle ratios, visualised on a
ternary diagram in which each point represents a distinct
anisotropic deployed configuration.

We scan the admissible $(\alpha_1, \alpha_2, \alpha_3)$
space and evaluate $\eta$ and $\epsilon_\text{bist}$ across
the full range $\beta \in [0,\,\pi/20]$
(Fig.~\ref{fig:ternary_plot}b), with representative slices
at $\beta = 0$ and $\beta = 0.13$~rad. The bistable region
forms a well-defined volume in the three-dimensional
$(\alpha_1, \alpha_2, \beta)$ space
(Figs.~\ref{fig:ternary_plot}c--f), confirming that
bistability is robustly maintained across a broad range of
anisotropic configurations and tilting angles. The bistable
region remains largely invariant in shape and location with
$\beta$, exhibiting a similar hexagonal admissible region
in both cases. Increasing $\beta$ produces two systematic
effects: a modest contraction of the admissible region
(reducing $\alpha_{\max}$ from $68^\circ$ to $65^\circ$),
and a more significant shift of $\epsilon_\text{bist}$
toward lower values across the entire admissible space
(from $0.29$--$0.43$ at $\beta = 0$ to $0.23$--$0.33$ at
$\beta = 0.13$~rad), while the overall range of $\eta$ is
largely preserved.

These observations establish $\beta$ as a tuning parameter
for the energy landscape: the set of anisotropic
deformation states admitting bistability is largely
preserved across $\beta$, while the strain at which the
bistable equilibrium occurs and the relative depth of the
energy well are systematically controlled. This is the key
design handle for the inverse design framework in
Section~\ref{sec:design}: given a target shape defined by
$(\alpha_1, \alpha_2, \alpha_3,\epsilon_\text{bist})$, $\beta$ is selected such
that deployed shape matches the geometric
compatibility requirement, ensuring the deployed
tessellation snaps into its target configuration as a
stable equilibrium.

\subsection{Mechanically-aware bistable design}
\label{sec:design}

\begin{figure}[ht!]
    \centering
    \includegraphics[width=0.92\linewidth]{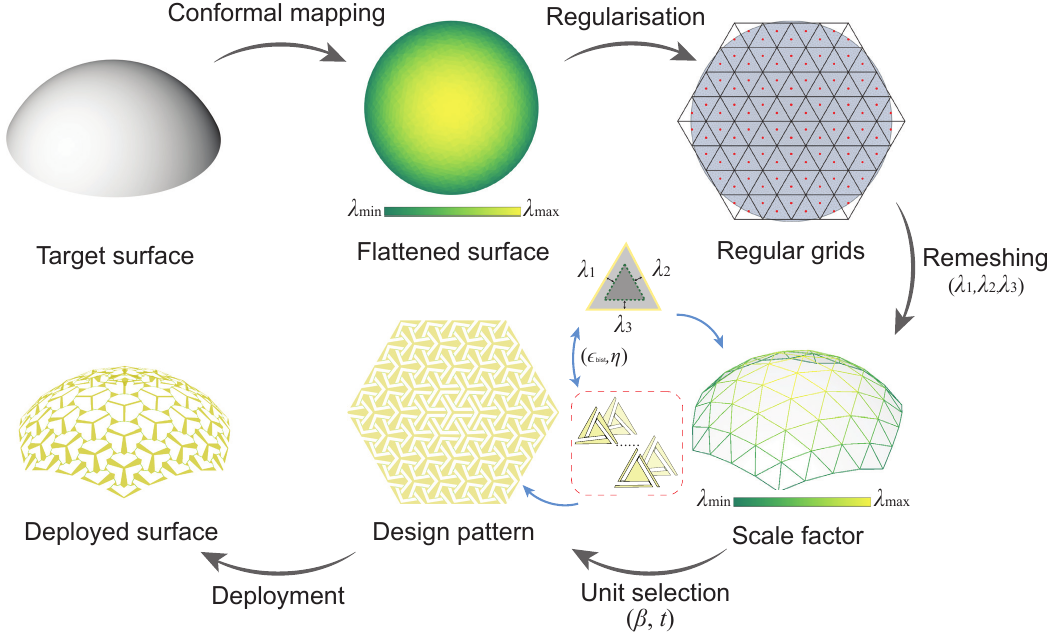}
    \caption{\textbf{Inverse design framework for bistable kirigami
    tessellations.}
    The target surface is conformally flattened via BFF, yielding a
    spatially varying scale-factor field $[\lambda_{\min},\lambda_{\max}]$.
    A regular triangular grid is overlaid on the planar domain; barycentric
    interpolation converts vertex-based scale factors into edge-wise values
    $(\lambda_1,\lambda_2,\lambda_3)$ for each element.
    Each element is then assigned a bistable unit from the precomputed
    library — parametrised by $(\beta,t)$ — that matches the required
    scale factors while maximising $\eta$.
    Assembling the selected units yields the flat cut pattern, which deploys onto the target surface.}
    \label{fig:workflow}
\end{figure}

Building on Section~\ref{sec:bistability}, we develop an inverse design
framework for programmable deployable surfaces composed of heterogeneous
bistable units (Fig.~\ref{fig:workflow}). The framework extends the
conformally transformable design procedure of~\cite{wang2023kirigami}---limited to isotropically deforming units---by incorporating anisotropic
deployment throughout. It consists of three stages: (1)~a parametric sweep
producing a precomputed bistable unit library (Section~\ref{sec:bistability});
(2)~surface parametrisation, which maps the target surface to a planar
domain and extracts edge-wise scale factors
(Section~\ref{sec:geometry}); and (3)~unit selection, which assigns an
optimal bistable unit to each grid element
(Section~\ref{sec:unit_selection}).

\subsubsection{Surface parametrisation}
\label{sec:geometry}

A conformal map $f:\mathcal{S}\to\mathbb{R}^2$ is constructed using
Boundary First Flattening (BFF)~\cite{sawhney2017boundary}, which
preserves local angles and admits direct boundary control. The pullback
metric satisfies $\tilde{g}=\lambda^2(\mathbf{x})\,g$, where
$\lambda(\mathbf{x})>0$ encodes the local isotropic stretch (Supplementary ~\ref{sec:S_conformal}); its spatial
variation reflects the Gaussian curvature of $\mathcal{S}$ and is
necessarily non-uniform for curved surfaces. Conformal flattening is
adopted because it distributes metric distortion isotropically at each
point, minimising finite-size anisotropy in the grid triangles for a
given resolution.

A regular triangular grid with side length $L$ is then overlaid on the
flattened domain. The scale factor at each grid node is obtained by
barycentric interpolation from surrounding mesh vertices, and converted
into edge-wise scale factors (Supplementary ~\ref{sec:S_regularisation})
\begin{equation}
    \lambda_i = \frac{\tilde{\ell}_i}{\ell_i}, \quad i=1,2,3,
    \label{eq:lambda_edge}
\end{equation}
where $\ell_i$ and $\tilde{\ell}_i$ are the undeformed and deformed edge
lengths. These three values fully characterise the local anisotropic
deformation of each grid element.

\subsubsection{Unit selection}
\label{sec:unit_selection}

The edge-wise scale factors $(\lambda_1,\lambda_2,\lambda_3)$ from
Eq.~(\ref{eq:lambda_edge}) are converted into deployed triangle geometry
$(\alpha_1,\alpha_2,\alpha_3,\epsilon_\text{bist})$, mapping each grid element into the anisotropic
deformation space characterised in Section~\ref{sec:bistability}. The
tilting angle $\beta$ is then selected from the precomputed library such
that the bistable equilibrium coincides with the target deployed
configuration while maximising $\eta$.

The ligament thickness $t$ is varied spatially as an independent design
handle: since $\Delta E\propto Et^3$, elements requiring higher
perturbation resistance are assigned greater $t$, while elements with
smaller deformation demands receive reduced $t$ to limit stress
concentration under repeated snap-through~\cite{shang2018durable}. The
slenderness ratio $t/l\leq 1/12$ is enforced throughout. The assembled
heterogeneous tessellation encodes the full surface curvature in its flat
cut pattern and, upon deployment, snaps bistably into the prescribed
three-dimensional configuration.

\subsubsection{Examples}
\label{sec:examples}

\begin{figure}[ht!]
    \centering
    \includegraphics[width=0.97\linewidth]{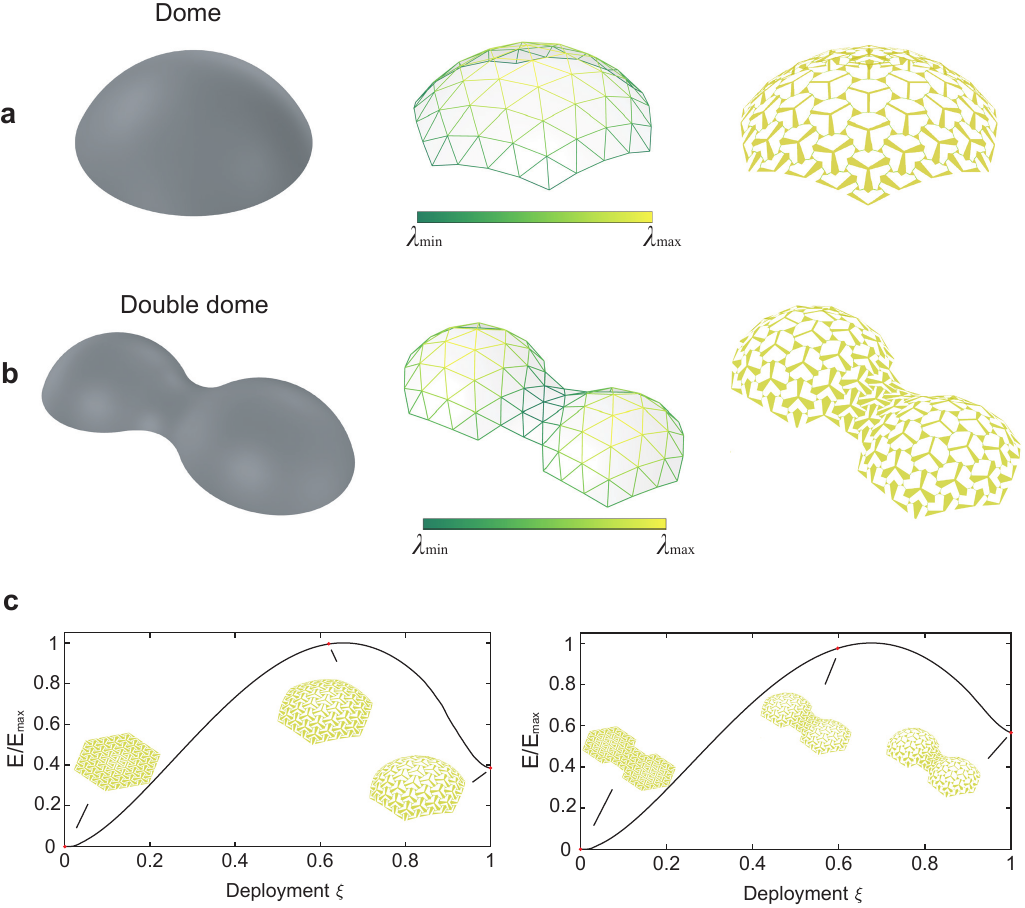}
    \caption{\textbf{Inverse design applied to two target surfaces.}
    (a)~Dome ($K>0$) and (b)~double dome (mixed $K$): target geometry,
    conformal scale-factor field, and the deployed tessellation configuration.
    (c) ~Normalised global in-plane deployment energy
    $E/E_{\max}$ versus deployment parameter $\xi$ for the dome (left) and double dome (right). }
    \label{fig:workflow_sample}
\end{figure}

Two surfaces illustrate the framework: the dome ($K>0$ throughout)
and the double dome (mixed $K$). The global deployment energy is
obtained by summing the in-plane ligament strain energy across all
units along the interpolated path (Fig.~\ref{fig:workflow_sample}).
Out-of-plane bending---arising from the dihedral angles between
adjacent faces---is excluded from the optimisation, but this
involves no loss of generality: once the target surface is determined,
the dihedral angles are fully determined by the geometry, so the
out-of-plane contribution does not affect the optimal unit selection. Both surfaces yield a well-defined bistable energy profile, confirming stable deployment under heterogeneous anisotropic deformation.

\subsection{Fabrication and Experiment}
\label{sec:experiment}

Physical specimens were fabricated by laser cutting 1.5~mm thick
Delrin Acetal sheets. Delrin was selected for its high stiffness and
elastic recovery, making it representative of rigid-material bistable
systems — the primary target application of the proposed framework.
The ligament slenderness ratio $t/l$ is assigned spatially, following
the unit selection procedure of Section~\ref{sec:unit_selection}, and
varies across the tessellation to balance bistability, durability, and
minimum feature size imposed by the laser cutting process.

A practical constraint arises at the boundary of the tessellation:
rim units lack the neighbouring cells that would otherwise impose the
geometric constraints required for deployment. To address this, the
conformal scale-factor field is globally rescaled so that rim units
retain $\lambda=1$ (remaining undeployed), while interior units are
adjusted accordingly. Because the deployed surface geometry depends
only on the relative distribution of scale factors, this rescaling
preserves the target shape while ensuring boundary compatibility.

\begin{figure}[!ht]
    \centering
    \includegraphics[width=\linewidth]{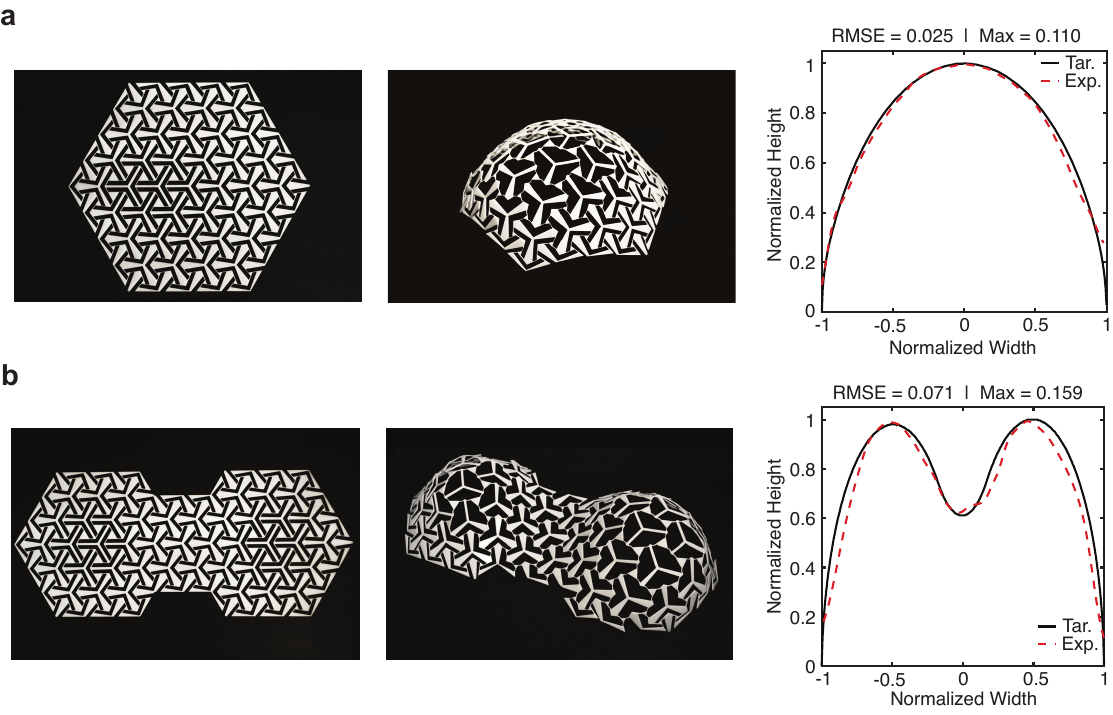}
    \caption{\textbf{Fabrication and experimental validation.}
    (a)~Dome and (b)~double dome: flat cut pattern (left), 
    free-standing deployed specimen (centre), and front-view 
    contour comparison between target and experiment (right). 
    RMSE and maximum deviation are reported as fractions of 
    the normalised span.}
    \label{fig:experiments}
\end{figure}

Shape accuracy was assessed by comparing the front-view contour of
the deployed specimen against the target profile
(Fig.~\ref{fig:experiments}). Photographs were taken with the sample
in its free-standing stable state, confirming bistable deployment
without external constraints. The dome achieves RMSE~$= 0.025$ with
a maximum deviation of~$0.110$; the double dome yields
RMSE~$= 0.071$ and maximum deviation~$0.159$. 

\section{Discussion and Conclusion}
To conclude, this work bridges the gap between rigid-material 
fabrication and programmable shape-morphing structures by combining 
a semi-analytical mechanical model, anisotropic deployment analysis, 
and conformal surface parametrisation into a unified inverse design 
framework for bistable kirigami tessellations. Unlike soft 
shape-morphing systems, which exhibit weak bistability and rely on 
external constraints or continuous actuation to maintain deployed 
configurations, the proposed framework exploits instability-induced 
snap-through in slender elastic ligaments to achieve self-locking 
deployment with a well-defined energy barrier---endowing the structure with inherent load-carrying capability in both stable states.

The proposed semi-analytical model greatly simplifies the 
optimisation process, accurately capturing the full energy profile 
of ligament-based kirigami structures without resorting to numerical methods. Building on this, we demonstrate that bistability 
is governed not by overall area expansion alone but by the specific 
anisotropic deployment path to capture accurate energy evolvement. These unit-level insights are 
embedded in a three-stage inverse design pipeline that translates an 
arbitrary target surface into a flat cut pattern, which deploys 
into the prescribed three-dimensional configuration without 
external constraints. Validation on dome and double-dome geometries, 
fabricated by laser cutting from rigid Delrin sheet, confirms stable 
deployed configurations with shape accuracies of RMSE~$= 0.025$ and 
RMSE~$= 0.071$, respectively.

Several limitations remain. The admissible scale-factor range of the 
triangular bistable unit is inherently bounded by the bistable 
region characterised in Section~\ref{sec:bistability}, which 
restricts the class of target surfaces that can be accurately 
approximated — highly complex geometries requiring large or strongly 
anisotropic local stretch may fall outside this range. Furthermore, 
the current framework optimises in-plane ligament energy only; 
out-of-plane bending at the dihedral interfaces between adjacent 
faces is not accounted for in the design stage. This omission 
introduces a bias towards surfaces of positive Gaussian curvature, 
where out-of-plane contributions are modest, while surfaces of 
negative curvature — such as saddle shapes — incur larger 
out-of-plane energy penalties that may suppress or eliminate 
bistability. Incorporating out-of-plane mechanics into the 
optimisation pipeline, alongside exploration of alternative unit 
cell topologies with wider admissible deformation ranges can be explored in the future.

\section*{Acknowledgments}

MAD acknowledges UKRI for support under the EPSRC Open Fellowship scheme (Project No. EP/W019450/1).

\bibliographystyle{unsrt}  
\bibliography{IAHR}

\clearpage

\section*{Supplementary Material}

\setcounter{figure}{0}
\renewcommand{\thefigure}{S\arabic{figure}}

\setcounter{table}{0}
\renewcommand{\thetable}{S\arabic{table}}

\setcounter{equation}{0}
\renewcommand{\theequation}{S\arabic{equation}}

\setcounter{section}{0}
\renewcommand{\thesection}{S\arabic{section}}

%
%
%
%


\section{Discrete Beam Model}
\label{sec:S_HBM}

\begin{figure}[ht!]
    \centering
    \includegraphics[width=0.93\linewidth]{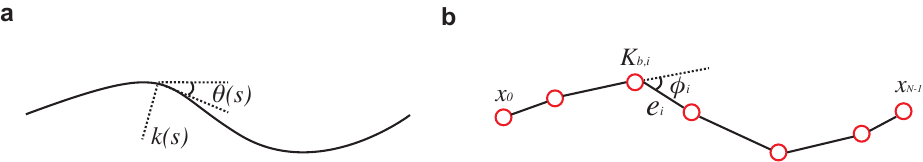}
    \caption{(a) Diagram of Euler beam in continuous case, and (b) Diagram of Euler beam in discrete case}
    \label{fig:hbm}
\end{figure}

We model each ligament using the Hencky bar-chain model (HBM), which
discretises the continuous Euler--Bernoulli beam into $N$ extensible
bars connected by $K = N-1$ rotational springs, allowing both bending
and axial deformations to be quantified. Consider an Euler--Bernoulli
beam of undeformed length $L$, bending rigidity $EI$, and axial
rigidity $EA$, parameterised by arc length $s \in [0, L]$. Let
$\theta(s)$ be the tangent angle and $\kappa(s) = \theta'(s)$ the
curvature. The axial strain is $\epsilon(s) = \lambda(s) - 1$, where
$\lambda(s)$ is the local stretch ratio. The total elastic energy is
\begin{equation}
    E_\text{tot} = \frac{1}{2} EI \int_0^L \kappa(s)^2 \,\mathrm{d}s
    + \frac{1}{2} EA \int_0^L \epsilon(s)^2 \,\mathrm{d}s.
    \label{eq:S_energy_continuous}
\end{equation}
The beam is approximated by $N$ bars of undeformed lengths $\{l_i\}_{i=1}^N$
($\sum_i l_i = L$) connected by $N-1$ hinges. The hinge rotation at
$x_i$ is $\phi_i = \theta_{i+1} - \theta_i$. The HBM assigns a
torsional spring at each hinge with stiffness
\begin{equation}
    K_{b,i} = \frac{EI}{h_i}, \qquad h_i \equiv \tfrac{1}{2}(\ell_i + \ell_{i+1}),
\end{equation}
and an axial spring on each bar with stiffness $K_{a,i} = EA/\ell_i$.
The discrete energy is
\begin{equation}
    E_b^{(N)} = \sum_{i=1}^{N-1} \tfrac{1}{2} K_{b,i} \phi_i^2, \qquad
    E_a^{(N)} = \sum_{i=1}^{N} \tfrac{1}{2} K_{a,i} e_i^2,
\end{equation}
so that $E_\text{tot}^{(N)} = E_b^{(N)} + E_a^{(N)}$.

The deformation is induced by prescribed rotation and translation at
the two ligament ends. Let $A, B \in \mathbb{R}^2$ be the end positions
and $\Theta_A$, $\Theta_B$ the prescribed end tangent angles. The
geometric compatibility conditions are
\begin{align}
    \text{(Angle constraint)} \quad &
    \Theta_A + \sum_{i=1}^{N-1} \phi_i = \Theta_B,
    \label{eq:S_angle} \\
    \text{(Position closure)} \quad &
    A + \sum_{i=1}^{N} (l_i + e_i)
    \begin{bmatrix} \cos\theta_i \\ \sin\theta_i \end{bmatrix} = B.
    \label{eq:S_position}
\end{align}
The equilibrium configuration minimises $E_\text{tot}^{(N)}$ subject
to Eqs.~(\ref{eq:S_angle})--(\ref{eq:S_position}), solved using the
interior-point method (\texttt{fmincon}, MATLAB).

\subsection*{Discretisation strategy and convergence}

\begin{figure}[ht!]
    \centering
    \includegraphics[width=0.93\linewidth]{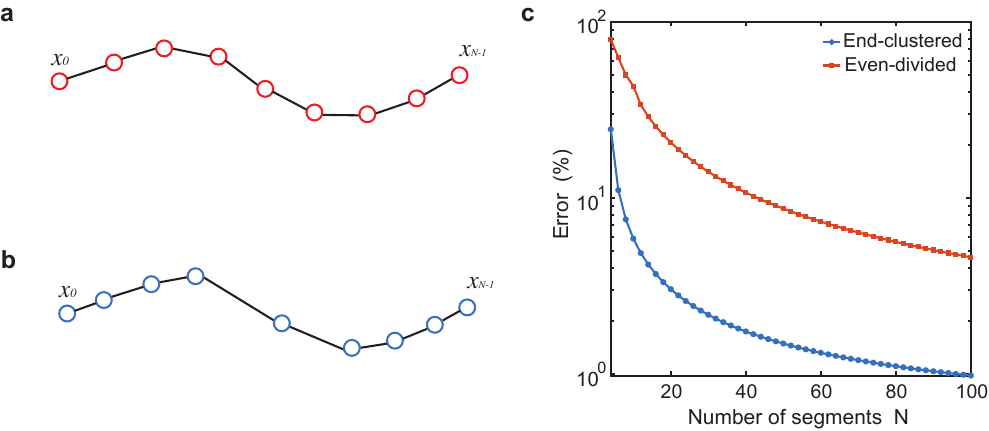}
    \caption{(a)~Even-divided HBM. (b)~End-clustered HBM.
    (c)~Convergence of relative energy error versus number of
    segments $N$ for both schemes; the end-clustered scheme reaches
    5\% error at $N=12$, compared with $N \approx 100$ for the
    even-divided scheme.}
    \label{fig:S_convergence}
\end{figure}

We compare two discretisation schemes for the Hencky bar-chain model.
In the even-divided scheme (Fig.~\ref{fig:S_convergence}(a)), the ligament
is partitioned into $N$ segments of equal length. In the end-clustered
scheme (Fig.~\ref{fig:S_convergence}(b)), the segment density is increased
towards the two ends, where the bending curvature---and hence the
strain-energy density---is largest; refining the discretisation in these
high-curvature regions resolves the rapid spatial variation of the solution
with far fewer degrees of freedom.

We quantify accuracy through the relative energy error against a converged
FEM reference,
\begin{equation}
    \text{Error}(N) = \frac{\bigl|E^{(N)} - E_\text{FEM}\bigr|}{E_\text{FEM}}.
\end{equation}
The even-divided scheme requires $N \approx 100$ segments to bring this error
below 5\%, whereas the end-clustered scheme attains the same accuracy with
only $N = 12$ segments (Fig.~\ref{fig:S_convergence}(c)). We therefore adopt
the end-clustered scheme throughout, as it preserves numerical accuracy while
substantially reducing the number of degrees of freedom.

\section{Isotropic Deployment}
\label{sec:S_isotropic}

\begin{figure}[ht!]
    \centering
    \includegraphics[width=0.90\linewidth]{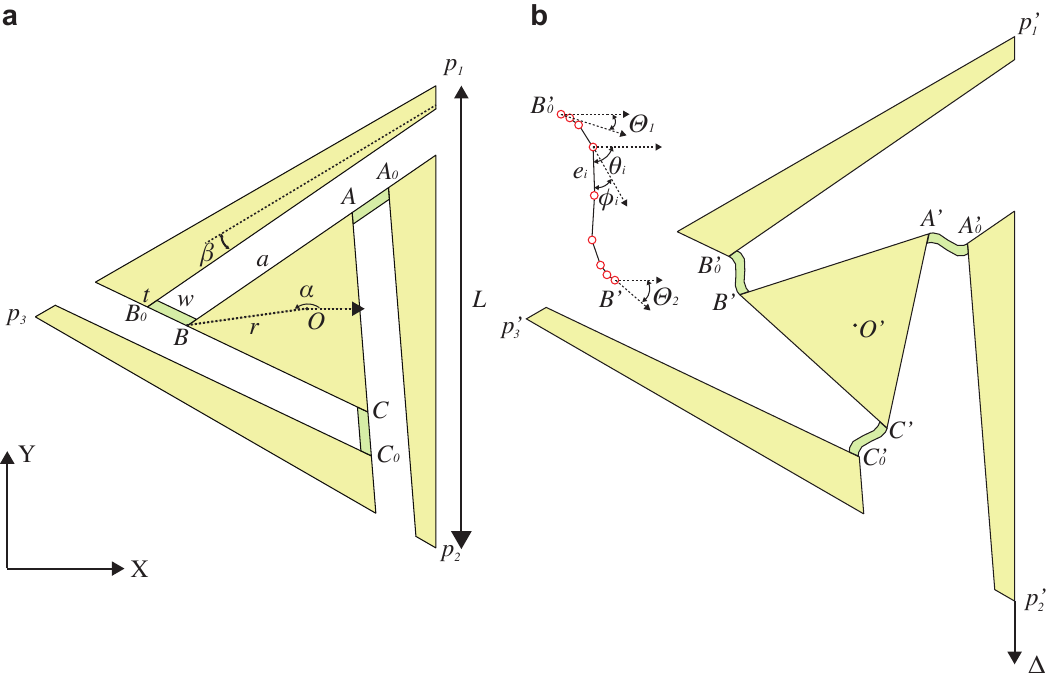}
    \caption{(a)~Original unit geometry. (b)~Deployed unit with applied
    displacement $\Delta$ on node $p_2$; node $p_1$ is fixed.}
    \label{fig:S_isotropic_unit}
\end{figure}

For an isotropically deploying unit, the three-fold rotational symmetry
of the triangular unit ensures all three ligaments undergo identical
deformation; it is therefore sufficient to analyse a single
representative ligament. The geometries of the original and deployed
units are shown in Fig.~\ref{fig:S_isotropic_unit}. The centroid of
the triangular core coincides with the centroid of the entire unit. The
left-end tangent angle is $\Theta_1 = \pi/3 - \beta$, and the
right-end tangent angle is determined by the vector $\overrightarrow{BC}$,
\begin{equation}
    \overrightarrow{BC} = \mathbf{C} - \mathbf{B} = \sqrt{3}\,r
    \begin{bmatrix}
        \cos\!\left(\alpha + \tfrac{5\pi}{6}\right) \\
        \sin\!\left(\alpha + \tfrac{5\pi}{6}\right)
    \end{bmatrix},
\end{equation}
giving $\Theta_2 = \alpha + 5\pi/6$. 

The equilibrium configuration of the ligament is obtained by minimising its total elastic energy---the sum of the bending energy stored in the rotational springs and the axial (stretching) energy stored in the bars---subject to two geometric compatibility conditions. The \emph{angle condition} $g_\Theta$ requires the rotations accumulated along the chain to reproduce the prescribed end tangent angles $\Theta_1$ and $\Theta_2$, which are fixed by the rigid connections of the ligament to the flank and to the triangular core. The
\emph{position closure condition} $g_r$ requires the deformed chain to span exactly the gap between its two attachment points, so that the ligament stays joined to the flank and core with no break.
\begin{equation}
\begin{aligned}
    \min_{\mathbf{x}} \quad & E_\text{tot}(\mathbf{x})
    = \tfrac{1}{2}\sum_{i=1}^{K} K_{b,i}\phi_i^2
    + \tfrac{1}{2}\sum_{i=1}^{N} K_{a,i} e_i^2, \\
    \text{s.t.} \quad &
    g_\Theta(\mathbf{x}) = \Theta_1 + \sum_{i=1}^{K}\phi_i - \Theta_2 = 0, \\
    & g_r(\mathbf{x}) = \mathbf{B}_0 + \sum_{i=1}^{N}(l_i+e_i)
    \begin{bmatrix}\cos\theta_i\\\sin\theta_i\end{bmatrix}
    - \mathbf{O} - r\begin{bmatrix}\cos\alpha\\\sin\alpha\end{bmatrix} = \mathbf{0},
\end{aligned}
\label{eq:S_isotropic_opt}
\end{equation}
where $\mathbf{x} = [\phi_1,\ldots,\phi_K,\, e_1,\ldots,e_N,\, \alpha]^\mathsf{T}$,
$K_{b,i} = EI/h_i$, $K_{a,i} = EA/\ell_i$, and $\theta_1 = \Theta_1$,
$\theta_{i+1} = \theta_i + \phi_i$. This non-linear constrained optimisation
problem is solved with \texttt{fmincon} (\textsc{MATLAB}) using the
interior-point algorithm, with analytically supplied gradients for both the
objective and the constraints. 

\begin{figure}[!ht]
    \centering
    \includegraphics[width=0.95\linewidth]{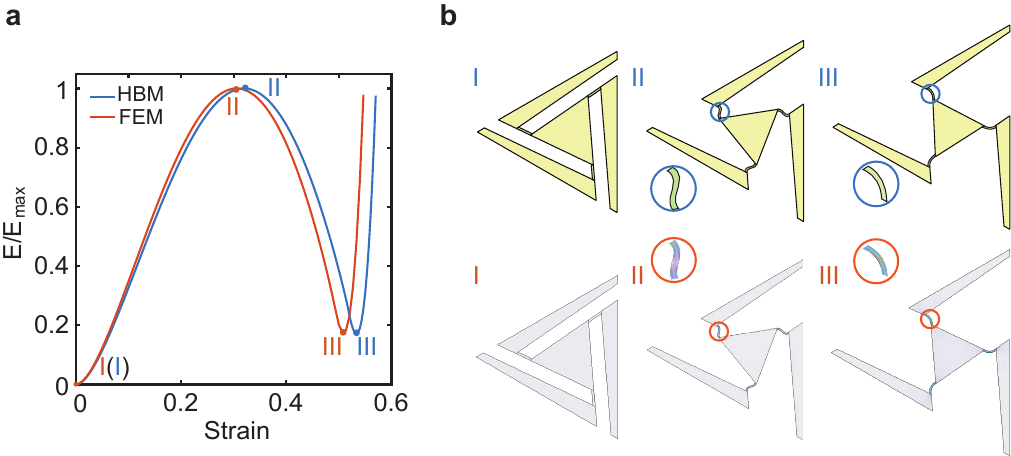}
    \caption{\textbf{Validation of the semi-analytical HBM model against FEM.}
    (a)~Normalised energy-strain response comparing HBM and FEM predictions.
    (b)~Deployment sequence predicted by HBM (top) and FEM (bottom) at 
    states~I--III.}
    \label{fig:deployment_energy}
\end{figure}

Figure~\ref{fig:deployment_energy} shows the deployment pathway and energy 
evolution of a representative auxetic unit ($w/L = 0.1$, $t/L = 0.015$, 
$a/L = 0.6$, $\beta = \pi/40$). Under isotropic deployment, the three-fold 
rotational symmetry of the triangular unit ensures all three ligaments 
undergo identical deformation; it is therefore sufficient to analyse a 
single representative ligament, with the total unit energy scaled by three.

The HBM and FEM results show good overall agreement across the full 
deployment range. The small strain offset at state~III arises from two 
modelling simplifications: the rigid-body assumption applied to the flanks and inner core, which neglects their elastic contributions, and the finite segment discretisation adopted in the HBM. Ligament snap-through --- maximum curvature at state~II transitioning to a stable configuration at state~III --- is accurately captured by both methods.

The energy profile in Fig.~\ref{fig:deployment_energy}a defines two key 
metrics used throughout this work. The \emph{bistability index} $\eta = 
\Delta E / E_{\max}$ quantifies the depth of the energy well, where $\Delta 
E = E_\text{II} - E_\text{III}$ and $E_{\max} = E_\text{II}$; a unit is 
bistable if and only if $\eta > 0$. The \emph{bistable strain} 
$\epsilon_\text{bist}$ is the strain at the local energy minimum, satisfying
\begin{equation}
    \left.\frac{\mathrm{d}E}{\mathrm{d}\epsilon}
    \right|_{\epsilon = \epsilon_\text{bist}} = 0,
    \qquad
    \left.\frac{\mathrm{d}^2 E}{\mathrm{d}\epsilon^2}
    \right|_{\epsilon = \epsilon_\text{bist}} > 0,
    \label{eq:bistable_strain_def}
\end{equation}
corresponding to state~III. 

\section{Anisotropic Deployment}
\label{sec:S_anisotropic}

\begin{figure}[ht!]
    \centering
    \includegraphics[width=0.85\linewidth]{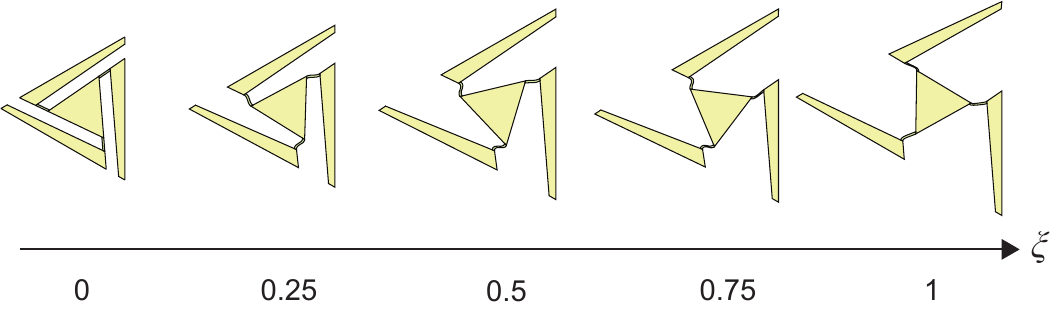}
    \caption{Anisotropic deployment of a single triangular kirigami
    unit at $\xi = 0,\, 0.25,\, 0.5,\, 0.75,\, 1$.}
    \label{fig:S_anisotropic_unit}
\end{figure}

For anisotropically deploying units, the three-fold symmetry no longer
holds, and all three ligaments must be analysed independently. The
deployment is parametrised by $\xi \in [0,1]$, with intermediate
nodal coordinates obtained by linear interpolation,
\begin{equation}
    \mathbf{q}(\xi) = (1-\xi)\,\mathbf{q}_0 + \xi\,\mathbf{q}_1,
\end{equation}
where $\mathbf{q}_0$ and $\mathbf{q}_1$ are the nodal coordinates of
the initial and fully deployed states, respectively. All units are
assumed to share the same $\xi$ at any point during deployment,
although their geometric changes differ. The anisotropic deployment
process of a single unit is illustrated in Fig.~\ref{fig:S_anisotropic_unit}.

Moreover, the centroid of the inner triangle is no longer coincident with the centroid of the unit, 
and its position must be treated as an additional unknown. 
The optimisation problem can be formulated as follows,

The full state vector collects unknowns from all three ligaments and
the inner triangle centroid,
\begin{equation}
    \mathbf{x} = \bigl[\boldsymbol{\phi}^{(1)},\, \mathbf{e}^{(1)},\,
    \boldsymbol{\phi}^{(2)},\, \mathbf{e}^{(2)},\,
    \boldsymbol{\phi}^{(3)},\, \mathbf{e}^{(3)},\,
    \alpha,\, x_G,\, y_G \bigr]^\mathsf{T},
\end{equation}
where $\boldsymbol{\phi}^{(\ell)}$ and $\mathbf{e}^{(\ell)}$ are the
hinge rotations and segment extensions of ligament $\ell$, and
$(x_G, y_G)$ is the centroid of the inner triangle. The total energy
sums contributions from all three ligaments,
\begin{equation}
    E_\text{tot}(\mathbf{x}) = \sum_{\ell=1}^{3}
    \left\{
    \frac{1}{2}\sum_{i=1}^{K} K_{b,i}^{(\ell)} \bigl(\phi_i^{(\ell)}\bigr)^2
    + \frac{1}{2}\sum_{i=1}^{N} K_{a,i}^{(\ell)} \bigl(\Delta e_i^{(\ell)}\bigr)^2
    \right\},
\end{equation}
where $\Delta e_i^{(\ell)} = e_i^{(\ell)} - \ell_i^{(\ell)}$,
$K_{b,i}^{(\ell)} = EI/h_i^{(\ell)}$, and
$K_{a,i}^{(\ell)} = EA/\ell_i^{(\ell)}$.
Geometric compatibility is enforced by angle and positional closure
constraints for each ligament $\ell = 1,2,3$,
\begin{align}
    g_\Theta^{(\ell)}(\mathbf{x}) &=
    \Theta_1^{(\ell)} + \sum_{i=1}^{K}\phi_i^{(\ell)}
    - \Theta_2^{(\ell)}(\alpha) = 0, \\
    \mathbf{g}_r^{(\ell)}(\mathbf{x}) &=
    \mathbf{B}_0^{(\ell)} + \sum_{i=1}^{N}
    \bigl(\ell_i^{(\ell)} + e_i^{(\ell)}\bigr)
    \begin{bmatrix}\cos\theta_i^{(\ell)}\\\sin\theta_i^{(\ell)}\end{bmatrix}
    - \mathbf{G} = \mathbf{0},
\end{align}
where $\mathbf{G} = [x_G, y_G]^\mathsf{T}$ is the centroid of the
inner triangle.

\section{Conformal Mapping and Discretisation}
\label{sec:S_conformal}

\subsection*{Smooth setting}

\begin{figure}[ht!]
    \centering
    \includegraphics[width=0.86\linewidth]{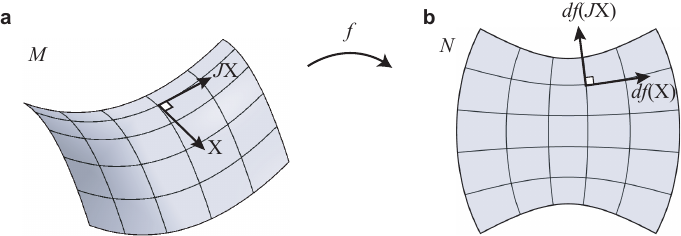}
    \caption{
    conformal mapping from (a) a smooth curved surface to (b) a flattened surface
    }
    \label{fig:smooth_setting}
\end{figure}

In the smooth setting, the objective is to construct a conformal map
$f:(M,g)\rightarrow(N,\tilde g)$ from a smooth surface $M$ with Riemannian
metric $g$ to a planar domain $N\subset\mathbb{C}$ with metric $\tilde g$,
such that the two metrics differ only by a positive scalar field. At each
point $\mathbf{x}\in M$ the pullback metric satisfies
\begin{equation}
\tilde g = \lambda^2(\mathbf{x})\, g,
\label{eq:conformal_metric}
\end{equation}
where $\lambda(\mathbf{x})>0$ is the local isotropic stretch (the ratio of
flattened to original length) and $\lambda^2$ the corresponding conformal
factor on the metric.

Under this conformal change, the Gaussian curvatures of the two metrics are
related by the Yamabe equation,
\begin{equation}
\tilde K = \frac{1}{\lambda^2}\bigl(K - \Delta_g \ln\lambda\bigr),
\label{eq:gaussian_curvature_conformal}
\end{equation}
where $\Delta_g$ is the Laplace--Beltrami operator on $(M,g)$, and $K$,
$\tilde K$ are the Gaussian curvatures of the original and target surfaces.
In the flattened coordinates the target metric is Euclidean, $\tilde g=\delta$,
so Eq.~\eqref{eq:conformal_metric} gives $g=\lambda^{-2}\delta$; the
Laplace--Beltrami operator of $g$ therefore reduces to a rescaled Euclidean
Laplacian,
\begin{equation}
\Delta_g = \lambda^{2}\,\Delta,
\label{eq:laplace_beltrami_conformal}
\end{equation}
with $\Delta$ the standard planar Laplacian. Flattening sets $\tilde K=0$, and
Eq.~\eqref{eq:gaussian_curvature_conformal} reduces to
\begin{equation}
K = \lambda^{2}\,\Delta \ln\lambda,
\label{eq:flat_curvature_relation}
\end{equation}
showing that the Gaussian curvature of the original surface is fully encoded
in the spatial variation of the conformal factor $\lambda$.

A classic example is the stereographic projection of a hemisphere onto the
plane. For a hemisphere of radius $R$,
\[
\mathbf{x}(\theta,\phi)=\bigl(R\sin\theta\cos\phi,\;R\sin\theta\sin\phi,\;R\cos\theta\bigr),
\qquad \theta\in[0,\tfrac{\pi}{2}],\ \phi\in[0,2\pi),
\]
the induced metric (with $u^1=\theta$, $u^2=\phi$ and $g_{ij}=\mathbf{x}_i\cdot\mathbf{x}_j$) is
\[
g = R^2\,\mathrm{d}\theta^2 + R^2\sin^2\theta\,\mathrm{d}\phi^2 .
\]
Projecting from the south pole via $r = R\tan(\theta/2)$ gives
\[
g=\left(\frac{2R^2}{R^2+r^2}\right)^{2}\bigl(\mathrm{d}r^2+r^2\,\mathrm{d}\phi^2\bigr)
=\left(\frac{2R^2}{R^2+r^2}\right)^{2}\tilde g ,
\]
so that, comparing with Eq.~\eqref{eq:conformal_metric},
\begin{equation}
\lambda(r)=\frac{R^2+r^2}{2R^2}.
\end{equation}
Substituting into Eq.~\eqref{eq:flat_curvature_relation} recovers the constant
curvature $K=1/R^2$, confirming the consistency of the formulation.

\subsection*{Discrete setting}
\label{conformal_mapping_discrete}
\begin{figure}[ht!]
    \centering
    \includegraphics[width=0.93\linewidth]{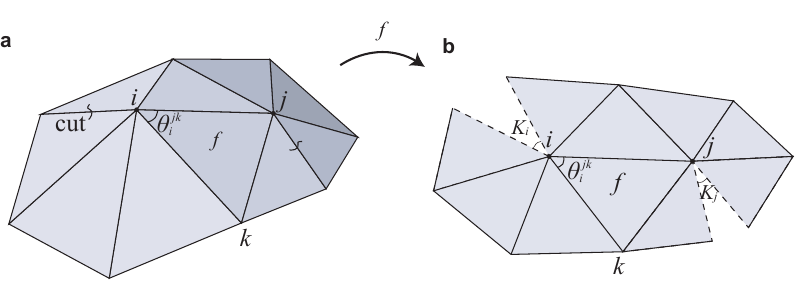}
    \caption{
    conformal mapping from (a) a discrete curved surface to (b) a flattened surface
    }
    \label{fig:discrete_setting}
\end{figure}

For a triangular mesh, exact conformality cannot be simultaneously
preserved at all elements. Gaussian curvature is concentrated at
vertices and defined by the angle deficit,
\begin{equation}
    K_i = 2\pi - \sum_{f_{ijk}\in F(i)} \theta_i^{jk},
\end{equation}
where $\theta_i^{jk}$ is the interior angle at vertex $i$ of face
$f_{ijk}$. A discrete conformal metric assigns a scale factor
$\lambda_i$ to each vertex; for an edge $(i,j)$ the scaled length is
\begin{equation}
    \tilde{\ell}_{ij} = \frac{\lambda_i + \lambda_j}{2}\,\ell_{ij}
    \equiv \lambda_{ij}\,\ell_{ij}.
\end{equation}
In practice, discrete conformal mappings preserve angles only in a
variational sense, so individual triangles may exhibit unequal edge
stretches, introducing locally anisotropic deformation.

\section{Regularisation}
\label{sec:S_regularisation}

To accommodate the designed unit geometry, a regular triangular grid
of equilateral elements with side length $L$ is overlaid on the
flattened mesh. Since the conformal scale factor is defined at mesh
vertices, the scale factor at each grid node $m$ within a mesh face
$(i,j,k)$ is obtained by barycentric interpolation,
\begin{equation}
    \lambda_m = w_i \lambda_i + w_j \lambda_j + w_k \lambda_k,
\end{equation}
where $(w_i, w_j, w_k)$ are the barycentric coordinates of $m$,
satisfying $w_i + w_j + w_k = 1$ and $w_i, w_j, w_k \geq 0$. The
vertex-based scale factors are then converted into edge-wise scale
factors $(\lambda_{mn}, \lambda_{np}, \lambda_{mp})$ for each grid
triangle $(m,n,p)$, which serve as input to the unit selection
procedure described in Section~2.3.2 of the main text.

\begin{figure}[ht!]
    \centering
    \includegraphics[width=0.85\linewidth]{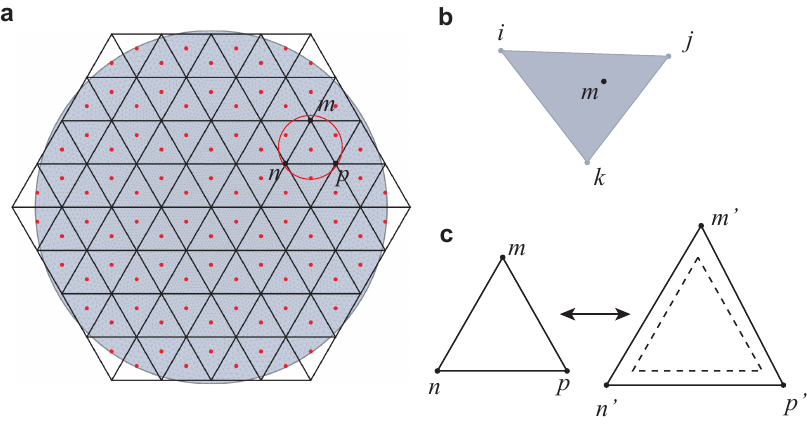}
    \caption{(a)~Regular triangular grid overlaid on the flattened
    mesh; red dots denote grid nodes whose scale factors are obtained
    by barycentric interpolation. (b)~Grid node $m^*$ enclosed within
    mesh face $(i,j,k)$. (c)~Anisotropic deployment from an
    equilateral grid triangle to a target triangle with unequal edge
    lengths.}
    \label{fig:S_regularisation}
\end{figure}

\begin{figure}[ht!]
    \centering
    \includegraphics[width=0.95\linewidth]{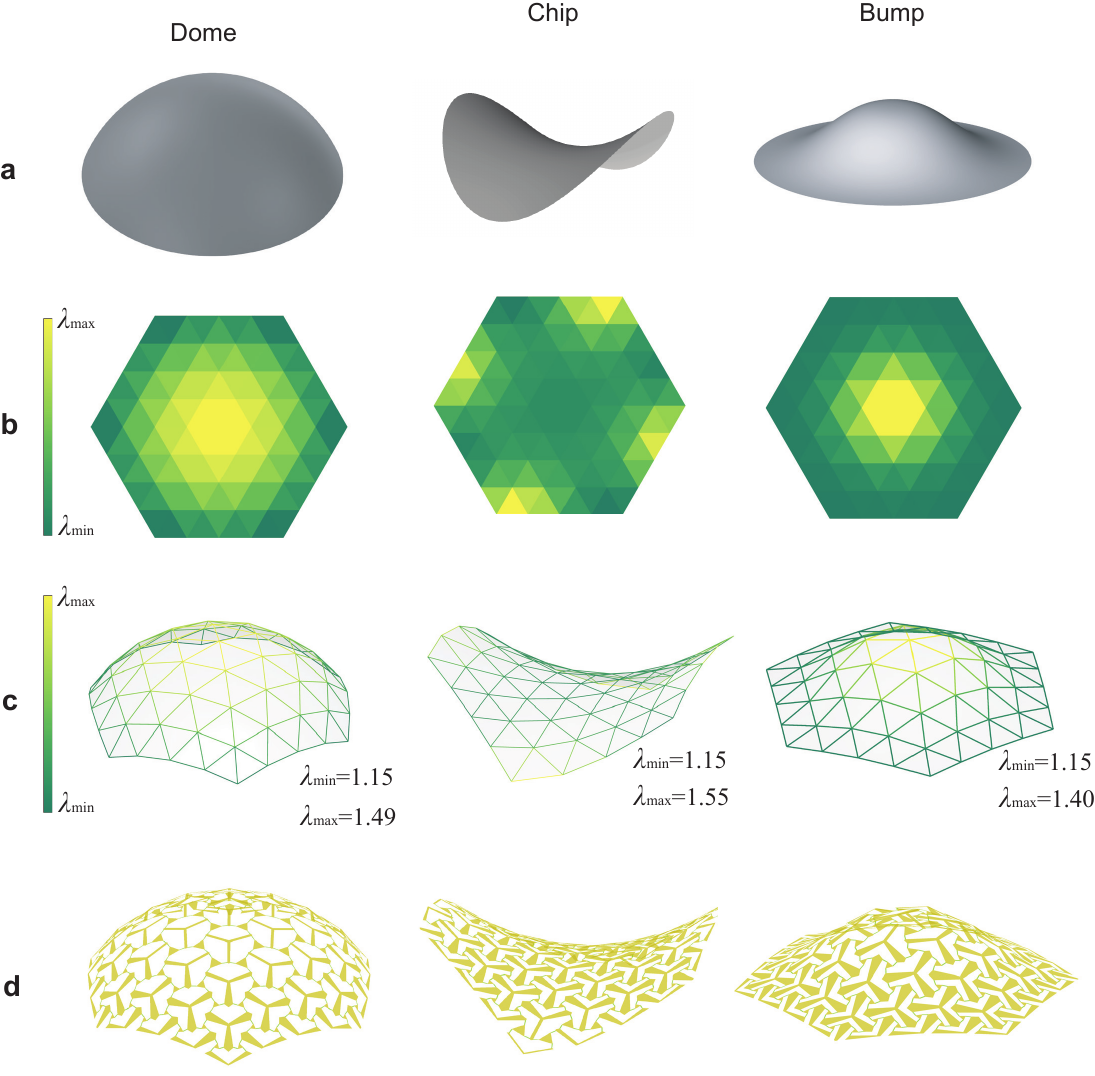}
    \caption{Geometric parametrisation for three target surfaces.
    (a)~Target surface. (b)~Discretised scale-factor field. (c)~Edge-wise scale factors. (d)~Tessellated kirigami sheet with optimised units}
    \label{fig:S_conformal_mapping}
\end{figure}

Figure~\ref{fig:S_conformal_mapping} illustrates the surface-parametrisation
pipeline for three target surfaces: the Dome, the Chip (a saddle), and the
Bump. For each case, conformal flattening (BFF) maps the target surface to
the plane and yields a spatially varying scale-factor field $\lambda(\mathbf{x})$
encoding the local stretch required to realise the surface curvature. This
field is regularised onto a regular triangular grid and reduced to the
edge-wise scale factors $(\lambda_1,\lambda_2,\lambda_3)$ that drive unit
selection.

%

\end{document}